\documentclass[%
 aip,
 amsmath,amssymb,
 reprint,%
]{revtex4-1}

\usepackage{graphicx}
\usepackage{dcolumn}
\usepackage{bm}

\usepackage[utf8]{inputenc}
\usepackage[T1]{fontenc}
\usepackage{mathptmx}

\begin{document}

\preprint{AIP/123-QED}

\title[]{Thermomagnetic instability of plasma composition gradients}

\author{James D. Sadler}
 \email{james4sadler@lanl.gov}
 \affiliation{%
Theoretical Division, Los Alamos National Laboratory, Los Alamos, NM 87545, USA}%
 \affiliation{%
Center for Nonlinear Studies, Los Alamos National Laboratory, Los Alamos, NM 87545, USA}%

\author {Hui Li}
 \affiliation{%
Theoretical Division, Los Alamos National Laboratory, Los Alamos, NM 87545, USA}%

\date{\today}

\begin{abstract}
We show that, under Braginskii magneto-hydrodynamics, anti-parallel gradients in average ion charge state and electron temperature can be unstable to the growth of self-generated magnetic fields. The instability is analogous to the field-generating thermomagnetic instability, although it is driven by the collisional thermal force magnetic source term rather than the Biermann battery term. The gradient in ion charge state causes a gradient in collisionality, which couples with temperature perturbations to create a self-generated magnetic field. This magnetic field deflects the electron heat flux in a way that reinforces the temperature perturbation. The derived linearized growth rate, typically on hydrodynamic  timescales, includes the resistive and thermal smoothing. It increases with large ion composition gradients and electron heat flux, conditions typical of the hohlraum walls or contaminant mix jets in inertial confinement fusion implosions. However, extended magneto-hydrodynamic simulations indicate that the instability is usually dominated and stabilized by the nonlinear Nernst advection, in a similar manner to the standard thermomagnetic instability.
\end{abstract}

\maketitle

%
\section{Introduction}
                       
Coupling of self-generated magnetic fields back into the fluid transport is a rich area of study in plasma dynamics. Many instabilities are known to arise, resulting in the spontaneous generation of magnetic fields in initially unmagnetized plasma. The energy source is often the fluid motion, or its higher order moments such as the electron heat flow and anisotropic pressure. For example, the Weibel instability was recently predicted \cite{PhysRevResearch.2.023080} to occur alongside laser plasma instabilities. Magnetized collisional transport is often used to explain the loss of symmetry at laser plasma ablation fronts \cite{colombant1977thermal, haines1986magnetic}, and was shown to be vital in correctly reproducing experiment \cite{gao2015precision}. Self-generated fields are also important in inertial confinement fusion fuel \cite{walsh2017self} and the surrounding hohlraum walls \cite{farmer2017simulation}. These processes were also the seed for magneto-genesis in the early universe \cite{tzeferacos2018laboratory, liao2019design, xu2009turbulence}, occurring even when the magnetic pressure is far less than the plasma pressure. 

The standard field generating thermomagnetic instability is one such example \cite{tidman1974field, bolshov1974spontaneous, 1950ZNatA...5...65B}. It causes exponential growth of transverse magnetic fields and temperature perturbations in non-uniform plasmas. It requires gradients in electron temperature $T_e$ and number density $n_e$. If there is a transverse temperature perturbation, Biermann battery magnetic fields will arise. The temperature gradient gives rise to a large heat flux which is deflected slightly by this magnetic field, in a direction that reinforces the temperature perturbation. 

In this work, we derive a related field generating instability that relies on gradients in ion charge state $Z$, rather than gradients in electron density. The magnetic field growth comes not from the collisionless Biermann mechanism, but from the collisional thermal force term. Gradients in the ion charge state lead to gradients in the local Coulomb collisionality and therefore the plasma transport coefficients. The growth rate, including various dissipation effects, is found to be similar to the standard field-generating instability \cite{tidman1974field}; it is on thermal transport and hydrodynamic timescales. 

Other terms in the extended magneto-hydrodynamics (ExMHD) equations act to dampen thermomagnetic instabilities. Thermal conduction will naturally smooth any transverse temperature perturbations, especially for small scale features. Plasma Ohmic resistance also dissipates the magnetic field. In addition, the resulting growth rates are often similar to hydrodynamic time-scales, meaning ion motion may be important. The necessary inclusion of ion fluid motion was found to significantly complicate the picture for the standard thermomagnetic instability \cite{ogasawara1980hydrodynamic}. Suppression also occurs due to kinetic non-local effects and electron gyro-radii exceeding the magnetic feature sizes \cite{sherlock2020suppression, PhysRevE.98.021201}. Finally, due to the large zeroth order temperature gradient, self-generated magnetic field will be rapidly Nernst advected into the cooler region \cite{hirao1981magnetic, colombant1977thermal, nishiguchi1985nernst}. Depending on the exact configuration, the linearized Nernst advection can dampen or enhance the instability. With ExMHD simulations of a localized higher-Z plasma region, we find that nonlinear Nernst advection is usually dominant and stabilizes the magnetic field growth. The developing field is advected into and then dissipated in the cooler and more resistive plasma, in a similar manner to the Nernst stabilization of the standard thermomagnetic instability \cite{sherlock2020suppression}.

In the second section of this work, we review the ExMHD model and the relevant transport effects. In the third section, we use this model, excluding Nernst advection, to derive a local linearized growth rate. In the fourth section, we conduct a global numerical simulation and find that the inclusion of Nernst advection serves to stabilize the Z-gradient mechanism. We summarize all of this in the fifth section.

\section{Extended MHD model}
In typical conditions, the plasma Debye scale $\lambda_D$ is much smaller than the density and temperature gradient scale-lengths $l_n$ and $l_T$. This means the plasma can be assumed quasi-neutral, such that the electron number density is $n_e=\sum_jZ_jn_j$, where $n_j$ and $Z_j$ are the number density and charge state of each ion species. If, in addition, the electron Coulomb mean free path is small compared to $l_n$ and $l_T$, the electric field is well represented by the extended magneto-hydrodynamics Ohm's law \cite{epperlein1986plasma}
\begin{align}
\begin{split}
   &\mathbf{E}= - \mathbf{u\times B}  + \frac{\mathbf{J\times\mathbf{B}}}{n_ee}-\frac{\nabla P_e}{n_ee}-\frac{\nabla.\underline\Pi_e}{n_ee}\\+ &\frac{m_e}{n_ee^2\tau}(\alpha_\perp\mathbf{J} + 
\alpha_\wedge\mathbf{J\times\hat b}) -  
   \frac{\beta_\perp}{e}\nabla T_e + \frac{\beta_\wedge}{e}\nabla T_e\times\mathbf{\hat b}. \label{ohm}
   \end{split}
\end{align}

In this equation, we have assumed a simplified two-dimensional geometry in which the magnetic field $\mathbf{B}$ is perpendicular to the Cartesian x-y plane and there are no gradients in the z direction. This is sufficient to explore the linear growth of the instability. Three-dimensional effects are left to future work. The electric field $\mathbf{E}$ is directed in the x-y plane and depends on the ion fluid velocity $\mathbf{u}$, the isotropic electron pressure $P_e=n_eT_e$, the traceless part of the electron pressure tensor $\underline\Pi_e$ and the current density $\mathbf{J}$. The magnetic field direction vector is $\mathbf{\hat b} = \mathbf{B}/|\mathbf{B}|$, along the positive or negative $z$ axis. 

Several electric field terms on the first line result from the relativistic transform from the electron fluid frame back to the laboratory frame. In addition, an electric field arises due to the electron pressure gradient. The terms on the second line of eqn. (\ref{ohm}) occur due to Coulomb collisions, and are thus dependent on the plasma transport coefficients \cite{epperlein1986plasma} $\alpha_\perp(\chi, \bar Z)$, $\alpha_\wedge(\chi, \bar Z)$, $\beta_\perp(\chi, \bar Z)$ and $\beta_\wedge(\chi, \bar Z)$. These are dimensionless functions of the local average ion charge state $\bar Z = (\sum_jn_jZ_j^2)/(\sum_jn_jZ_j)$ and the electron magnetization $\chi=e|\mathbf{B}|\tau/m_e$, where the electron Coulomb collision time
\begin{align}
    \tau&=\frac{3\sqrt{\pi}}{4}\frac{4\pi\epsilon_0^2m_e^2}{n_e\bar Z e^4 \ln(\Lambda)}\left(\frac{2T_e}{m_e}\right)^{3/2}\label{tau}\\
   &= \frac{3.4\times 10^5}{\bar Z\ln(\Lambda)} \left(\frac{T_\mathrm{e}}{\mathrm{eV}}\right)^{3/2}\left(\frac{n_\mathrm{e}}{\mathrm{cm}^{-3}}\right)^{-1} \,\,\,\mathrm{s},\label{tau2}\\
      \chi&= \frac{6.1\times 10^{16}}{\bar Z\ln(\Lambda)} \left(\frac{T_\mathrm{e}}{\mathrm{eV}}\right)^{3/2}\left(\frac{n_\mathrm{e}}{\mathrm{cm}^{-3}}\right)^{-1}\left(\frac{|\mathbf{B}|}{\mathrm{T}}\right).\label{chi2}
\end{align}
In these equations, $m_e$ is the electron mass, $e$ is the elementary charge and $\epsilon_0$ is the vacuum permittivity. The Coulomb logarithm is $\ln(\Lambda)$, where $\Lambda=4\pi n_e\lambda_D^3/\bar Z$. Eqn. (\ref{tau2}) gives a simple formula for $\tau$ in seconds, in terms of the electron temperature in electron-volts and electron number density per cm$^3$. Eqn. (\ref{chi2}) gives a simple formula for the dimensionless magnetization $\chi$ in terms of $|\mathbf{B}|$ in Tesla.

The magnetization $\chi$ is equivalent to the ratio of the electron Coulomb mean free path $\lambda_\mathrm{mfp}=\tau v_\mathrm{th}$ to its gyroradius, where $v_\mathrm{th}=\sqrt{2T_e/m_e}$ is the electron thermal speed. As such, when $\chi$ approaches one, the gyromotion becomes comparable to the Coulomb collisions, causing deflection and reduction of the current and heat flux. 

The $\alpha$ resistive terms in eq. (\ref{ohm}) occur due to Ohmic resistance from Coulomb collisions. Even when the current $\mathbf{J}$ is zero, faster electrons from the hotter region of a temperature gradient are less collisional [eq. (\ref{tau})], and so there is still a net force on electrons towards the colder region. The Ohm's law eqn. (\ref{ohm}) also therefore contains the $\beta_\perp$ and $\beta_\wedge$ collisional thermal force terms, acting on gradients in temperature $\nabla T_e$. The relative importance of electron-electron and electron-ion collisions depends on the ion charge state, leading to the $\bar Z$ dependence of the transport coefficients.

In a similar manner, the intrinsic electron heat flux $\mathbf{q}_e$ is also deflected by the magnetic field. Furthermore, heat flux can be driven by electric currents, as well as temperature gradients. Using the dimensionless $\beta$ and $\kappa$ transport coefficients \cite{epperlein1986plasma}, this results in the heat flux
\begin{align}
\mathbf{q}_e = &-\frac{n_eT_e\tau}{m_e}(\kappa_\perp\nabla T_e + \kappa_\wedge\mathbf{\hat b}\times \nabla T_e) - \frac{T_e}{e}(\beta_\perp\mathbf{J} + \beta_\wedge\mathbf{\hat b\times J}).\label{heat}
\end{align}
The magnetic field evolution is found by substitution of eqn. (\ref{ohm}) into Faraday's law, yielding the two-dimensional induction equation \cite{haines1986magnetic, walsh2020extended, sadler2020conference}
\begin{align}
\begin{split}
   \frac{\partial\mathbf{B}}{\partial t} =\,&-\nabla\times\mathbf{E}\\ =\,&\nabla\times\mathbf{(u_B\times B)} + D_R\nabla^2\mathbf{B} - \nabla D_R\times(\nabla\times\mathbf{B})\\ &+ \nabla\times\left(\frac{\nabla P_e + \nabla.\underline\Pi_e}{n_ee}\right) + \frac{1}{e}\nabla\beta_\perp\times\nabla T_e.\label{induction}
   \end{split}
\end{align}
This result has been simplified using several vector identities, as well as the magneto-hydrodynamics assumption of low frequency oscillations, such that $\mathbf{J}=c^2\epsilon_0\mathbf{\nabla\times B}$. In addition, the resulting terms that contain $\nabla.\mathbf{B}$ and $\nabla\times\nabla T_e$ are both zero. The magnetic diffusivity is given by $D_R=m_ec^2\epsilon_0\alpha_\perp/(n_ee^2\tau)$. 

The first term in eq. (\ref{induction}) advects the magnetic field with a velocity
\begin{align}
\begin{split}
\mathbf{u_B=u}-\frac{\mathbf{J}}{n_ee}\left(1+\frac{\alpha_\wedge}{\chi}\right) - \frac{\tau\beta_\wedge}{m_e\chi}\nabla T_e,\label{ub}
   \end{split}
\end{align}
composed of the ion fluid motion, the corrected Hall velocity and the Nernst advection down the temperature gradient.

There are several known plasma instabilities resulting from each of the terms in the induction equation eqn. (\ref{induction}). Magnetic amplification by dynamo action \cite{tzeferacos2018laboratory} depends on the fluid motion $\mathbf{u}$ in eqn. (\ref{ub}). The standard field generating thermomagnetic instability \cite{tidman1974field} results from coupling of the Biermann battery $\nabla P_e$ source term \cite{pert1977self} in eqn. (\ref{induction}) with the deflected Righi-Leduc $\kappa_\wedge$ heat flow in eqn. (\ref{heat}). The anisotropic pressure $\underline\Pi_e$ can generate magnetic fields that lead to the Weibel instability \cite{weibel}. Gradients in the $D_R$ plasma resistivity term can arise due to temperature variations, leading to the thermal instability \cite{haines1981thermal}. The resistive term is also unstable in the presence of a free-streaming charged beam in a pre-magnetized plasma \cite{bell2020instability}. Many of these instabilities are damped by the resistive diffusion of magnetic field given by $D_R\nabla^2\mathbf{B}$.

Finally, eqn. (\ref{ub}) shows that the $\beta_\wedge$ collisional thermal force term leads to advection of the magnetic field down temperature gradients \cite{colombant1977thermal}. This is known as Nernst advection. Nernst compression of a pre-existing zeroth-order magnetic field was also shown \cite{bissell2010field} to be unstable when combined with the deflected heat flow in eqn. (\ref{heat}). 

It was first noted by M. G. Haines \cite{haines1997saturation, haines1986heat} that the other consequence of the thermal force is the appearance of an additional magnetic field source term, the final term in eqn. (\ref{induction}). Since $\beta_\perp$ also has a $\bar Z$ dependence, this term is active even when $\mathbf{B}=0$. Although there are many subtle effects arising from plasma composition gradients\cite{yin2016plasma, PhysRevE.102.013212}, magnetic instabilities have barely been explored, since the $\bar Z$ dependence of $\beta_\perp$ is often neglected. Further study \cite{sadler2020conference, sadler2020magnetization} has found that in low $\bar Z$ plasma regions with steep gradients in $\bar Z$, this thermal force source term is actually of similar magnitude to the Biermann battery source term. This was previously verified via a kinetic simulation \cite{sadler2020magnetization}.

In this work, we go further and show that the $\nabla\beta_\perp\times\nabla T_e$ thermal force source term can become unstable. The instability occurs due to the resulting self-generated magnetic field coupling with the magnetized deflection of the heat flow given by eqn. (\ref{heat}). However, numerical simulations then show that, as in the case of the standard thermomagnetic instability \cite{sherlock2020suppression}, the Z-gradient instability is usually stabilized by the Nernst advection term in eqn. (\ref{ub}).

The instability is distinct from the field compressing process of Bissell \emph{et al.} \cite{bissell2010field}, in that it requires a gradient in $\bar Z$ rather than a pre-imposed magnetic field. It shares a closer resemblance to the field generating thermomagnetic instability \cite{tidman1974field}, although the field arises from the collisional thermal force source term and $\nabla \bar Z$, rather than from the Biermann source term and $\nabla n_e$. To verify this, we conduct the numerical simulation with uniform electron density.

Gradients in $\bar Z$ can occur in many situations in astrophysical and high energy density plasmas. In the latter, experimental targets are often composed of many different materials. Composition gradients may occur during laser ablation of hohlraum walls, or high-Z contaminant jets that can enter hydrogen fusion fuel. Even in plasma composed of a single element, there can be a $\nabla \bar Z$ if different regions have different ionization states. We assume that the ionization state is prescribed in the plasma by a detailed atomic physics model. 

Due to the $Z_j^2$ dependence of the Coulomb collision rates, these gradients in $\bar Z$ will change the relative importance of electron-electron and electron-ion collisions. This manifests itself in the $\bar Z$ dependence of the classical plasma transport coefficients \cite{epperlein1986plasma}. Gradients in the $\beta_\perp$ thermal force coefficient will then yield self-generated magnetic fields through the last term in eqn. (\ref{induction}), which couple back to the heat transport through eqn. (\ref{heat}).

\section{Growth Rate Derivation}

The linearized analysis proceeds as follows. Zeroth order quantities are denoted with subscript $0$ and small perturbed quantities will be denoted with subscript $1$. We assume there is no magnetic field at zeroth order. The plasma is assumed to start with zeroth order steady-state gradients in electron number density $n_e$, electron temperature $T_e$ and ion charge state $\bar  Z$, all directed along the $x$ direction. The gradient scale-lengths are $l_n = n_e/(\partial n_e/\partial x)$, $l_T = T_e/(\partial T_e/\partial x)$ and $l_Z = \bar Z/(\partial \bar Z/\partial x)$. Note that these lengths can also take negative values. We assume that the instability arises from first order transverse temperature and field perturbations $T_e(t, x, y) = T_0(x)+T_1(x, y,t)$ and $\mathbf{B} = B_1(x, y,t)\mathbf{\hat z}$, with $T_1, B_1\propto \exp(iky + \gamma t)$. The small magnetic field is assumed to result in $\chi\ll 1$. All other quantities are taken to have a zeroth-order dependence only along the $\mathbf{\hat x}$ direction. We also simplify the analysis by neglecting fluid motion ($\mathbf{u}=0$) and anisotropic pressure ($\underline\Pi_e=0$). 

It should be noted that the form of eqn. (\ref{induction}) is strictly rigorous only for single species plasma. Introduction of inter-species ion diffusion in the $x$ direction, where different ion species have different fluid velocities, yields additional terms in the ExMHD Ohm's law. These can be estimated using the results of reference \cite{molvig2014classical}. This gives an additional ion resistive term $|\mathbf{E}|\simeq  m_e\delta u_i/(e\tau)$ in the Ohm's law, where the relative ion diffusion velocity is approximately $\delta u_i\simeq\tau T_i/(l_Z\sqrt{m_im_e})$, $m_i$ is the light ion mass and $T_i$ is the ion temperature. In the case that $|l_T|\simeq |l_Z|$, this simplifies to $|\mathbf{E}|\simeq \sqrt{m_e/m_i}T_e/(el_T)$. This means the additional ion resistive term is smaller than the thermoelectric term by a factor of the square-root of the ion-electron mass ratio. However, reference \cite{molvig2014classical} derived this result in the ambipolar limit with zero magnetic field. Inclusion of our linearized magnetic field may lead to additional important ion transport terms\cite{zhdanov2002transport}, which we leave to future work. 

The results of this work are therefore only strictly valid for plasma with shallow composition gradients, such that inter-species diffusion is negligible. As a result, $n_e=n_0(x)$ and $\bar Z=\bar Z(x)$ are constant in time. This assumption will be further examined later in this section. Using the pressure $P_e=n_eT_e$, along with the two dimensional geometry and out of plane magnetic field, vector identities simplify equation (\ref{induction}) to
\begin{align}
\begin{split}
  \frac{\partial B_1}{\partial t} =\,&-\nabla.\left(\mathbf{u_B}B_1\right) + D_R\nabla^2B_1 \\&+ \nabla D_R.\nabla B_1 + \frac{\mathbf{\hat z}}{e}.\nabla( \beta_\perp - \ln (n_e))\times\nabla T_e.\label{inductionscalar}
   \end{split}
\end{align}

 The $\alpha_\wedge$, $\beta_\wedge$ and $\kappa_\wedge$ transport coefficients can be expanded, assuming $\chi\ll 1$, to give a linearized form  $\beta_\wedge(\chi, \bar Z)\simeq\chi\beta_{0\wedge}(\bar Z)$, where $\alpha_{0\wedge}(\bar Z)$, $\beta_{0\wedge}(\bar Z)$ and $\kappa_{0\wedge}(\bar Z)$ are given in reference \cite{epperlein1986plasma}. Meanwhile, the $\alpha_\perp$ and $\beta_\perp$ coefficients can be approximated, correct to second order, as their unmagnetized values $\alpha_\parallel(\bar Z) = \alpha_\perp(0,\bar Z)$ and $\beta_\parallel(\bar Z) = \beta_\perp(0,\bar Z)$. On substitution of $\mathbf{u_B}$ and the linearized forms, while neglecting terms quadratic in the small quantities, eqn. (\ref{inductionscalar}) simplifies to
\begin{align}
\begin{split}
  &\frac{\partial B_1}{\partial t} =\,\nabla.\left( B_1\frac{\mathbf{J}}{n_ee}(1+\alpha_{0\wedge}) + B_1\frac{\tau}{m_e}\beta_{0\wedge}\nabla T_0\right) \\&+ D_R\nabla^2B_1 + \frac{\partial D_R}{\partial x}\frac{\partial B_1}{\partial x} + \left(\frac{d\beta_\parallel}{d\bar Z}\frac{\bar Z}{l_Z} - \frac{1}{l_n}\right)\frac{ikT_1}{e}.\label{inductionscalar2}
   \end{split}
\end{align}
 The magnetic field is advected by the conservative advection term on the first line of eqn. (\ref{inductionscalar2}). Fluid motion is assumed to be zero, so the field is only advected by the Hall ($\mathbf{J}$) and Nernst ($\nabla T_e$) terms. The second line contains the $D_R\nabla^2 B_1$ diffusion term, the resistivity gradient term and the source term. This source term is comprised of the usual Biermann source containing $l_n$, as well as the $\bar Z$ gradient term. 
 
 With use of $\mathbf{J} = c^2\epsilon_0\nabla\times\mathbf{B}$ and $D_R = m_ec^2\epsilon_0\alpha_\parallel/(n_ee^2\tau)$, the ratio of the Hall and resistive terms is approximately
 \begin{align}
     \frac{|\nabla . (B_1\mathbf{J}(1+\alpha_{0\wedge})/(n_ee))|}{|D_R\nabla^2B_1|}\simeq \frac{eB_1\tau}{m_e} = \chi.
 \end{align}
 Our assumption of weak magnetization $\chi\ll 1$ therefore means the Hall advection is second order in $B_1$ and can be neglected with respect to the resistive diffusion. 
 
 The induction equation (\ref{inductionscalar2}) therefore simplifies to
\begin{align}
  \frac{\partial B_1}{\partial t} &= N + D_R\nabla^2B_1 + \frac{\partial D_R}{\partial x}\frac{\partial B_1}{\partial x} + \left(\frac{d \beta_\parallel}{d \bar Z}\frac{\bar Z}{l_Z} - \frac{1}{L_n}\right)\frac{ikT_1}{e},\label{inductionscalar3}\\
  N &= \frac{\partial}{\partial x}\left( B_1\frac{\tau}{m_e}\beta_{0\wedge}\frac{dT_0}{dx}\right).\label{Nernst}
\end{align}
So long as all quantities have a similar gradient scale length $|l_T|\simeq |l_n|\simeq |l_B|\simeq L$ in the $x$ direction, the ratio of the Nernst and resistivity gradient terms in eqn. (\ref{inductionscalar3}) is 
 \begin{align}
     \frac{|B_1\tau\beta_{0\wedge}T_0/(m_eL^2)|}{|D_RB_1/L^2|}\simeq \frac{T_0n_0e^2\tau^2}{m_e^2c^2\epsilon_0}=\ \frac{9\pi T_0}{2m_ec^2} \left(\frac{\Lambda}{\ln(\Lambda)}\right)^2 \equiv \Gamma^2.
 \end{align}
In a hot, weakly coupled plasma, the Nernst term is therefore of much greater magnitude than the resistivity gradient term. To derive a simple linearized growth rate, we therefore specialize to the case of $\chi\ll 1$ and $\Gamma \gg 1$ to neglect the $\partial D_R/\partial x$ term. The assumption $\Gamma\gg 1$ covers cases such as coronal plasma at a laser ablation front or contaminant jets within inertial confinement fusion fuel. 

We may now examine the condition for ion diffusion to be negligible. Since the Hall term is already neglected, if the inter-species ion diffusion velocity is small compared to the Hall velocity, the ion transport effects should also be negligible. This requires $\delta u_i\ll|\mathbf{J}|/(n_ee)$, equivalent to the condition $\chi kL\gg \Gamma^2\sqrt{m_e/m_i}$. Since we have assumed $\chi\ll 1$ and $\Gamma\gg 1$, this implies that $kL\gg 1$, known as the local approximation. For instability at lower transverse wavenumbers, the additional ion transport terms\cite{zhdanov2002transport} may be required. This is left to future work. We also note that since $kL\gg 1$, the $D_R\nabla^2B_1$ diffusive term, which contains the transverse $y$ derivative, may be of much larger magnitude than the $\nabla D_R$ term and it must be retained.

These considerations show that, for the assumed plasma conditions, the Nernst term $N$ has the largest magnitude and it cannot be neglected relative to the source term. Depending on the exact nonlinear temperature profile $T_0$, it can be positive or negative, either aiding or mitigating the thermomagnetic growth rate. At this point, some authors \cite{hirao1981magnetic} have neglected the $x$ dependence of the magnetic field $B_1$ and so removed $B_1$ from the derivative in eqn. (\ref{Nernst}). This means the Nernst process can be kept in the resulting final dispersion relation. However, neglecting the $x$ dependence of $B_1$ means that the Nernst advection term is not globally conservative of magnetic field, as an advection process should be. The spatial variation of the instability growth rate will naturally lead to $x$ dependence of $B_1$, invalidating this assumption. For a proper treatment of the Nernst advection, it is therefore necessary to retain the $x$ dependence of the perturbed quantities, requiring a global two-dimensional $x$-$y$ analysis \cite{bissell2015nernst}. This does not lead to a simple closed form linearized growth rate. To proceed further, we must therefore neglect the $x$ dependence of $B_1$ and, since this leads to an inconsistent and non-conservative Nernst term $N$, the derived growth rate cannot include the Nernst effect. Bissell \emph{et al.} have discussed these issues at length for the standard thermomagnetic instability \cite{bissell2013super, bissell2015nernst}.

We now find that the remaining terms become unstable in a local analysis, where the $x$ dependence of $B_1$ and $T_1$ are neglected. However, when the Nernst advection is later included in a global two-dimensional ExMHD simulation, we find that this Nernst advection stabilizes the process. 

To find a local dispersion relation, we therefore neglect $N$ and neglect the $x$ dependence of the growing wave-like perturbations to give $B_1(y,t), T_1(y,t)\propto \exp(iky+\gamma t)$. This results in
\begin{align}
\begin{split}
   &(\gamma + D_R k^2) B_1 = \left(\frac{d\beta_\parallel}{d\bar Z}\frac{\bar Z}{l_Z} - \frac{1}{l_n}\right)\frac{ikT_1}{e}.\label{B1}
   \end{split}
\end{align}

The magnetic field perturbation growth is tempered by the diffusion of magnetic field due to the Coulomb resistance term containing $D_R$.  The source term on the right hand side is a combination of the Biermann growth and an additional thermoelectric term due to the zeroth-order gradients in ion charge state $\bar Z$. The transport coefficient is approximately \cite{molvig2014classical}
\begin{align}
\beta_\parallel(\bar Z)\simeq \frac{30\bar Z (15\sqrt{2}+ 11\bar Z)}{288+604\sqrt{2}\bar Z + 217\bar Z^2}.
\end{align}

The derivative $d\beta_\parallel/d\bar Z$ is positive and is maximal for low $\bar Z$ plasmas, reaching a value of around $0.3$ at $\bar Z=1$. This means that, for low $\bar Z$ plasmas with $l_Z\simeq l_n$, the thermoelectric source term can be of similar magnitude to the Biermann source term. 

To proceed further, we must obtain an equivalent expression for $T_1$ from the linearized electron fluid energy equation
\begin{align}
&\frac{3n_0}{2}\frac{\partial T_1}{\partial t} = \nabla.\left(\frac{5}{2e}T_e\mathbf{J}-\mathbf{q}_e\right) + \mathbf{J.E} +S.\label{energyeqn}
\end{align}
This contains the divergence of the total heat flux, the Ohmic heating and a general source term $S(x)$, which is assumed to occur due to processes such as laser energy deposition, fusion reactions or radiative losses. Terms quadratic in $\mathbf{J}$ have been neglected.

Under our local assumption that $\partial B_1/\partial x=0$, this yields $\mathbf{J}= c^2\epsilon_0\nabla\times\mathbf{B}= c^2\epsilon_0ikB_1\mathbf{\hat x}$. With use of eqn. (\ref{ohm}), and again neglecting terms that are quadratic in the small quantities, the linearized $\mathbf{J.E}$ Ohmic heating term is
\begin{align}
\begin{split}
\mathbf{J.E} &\simeq \mathbf{J}.\left(-\frac{1}{e}(1+\beta_\perp)\nabla T_0 - \frac{T_0}{n_0e}\nabla n_0\right),\\
&= -i\frac{c^2\epsilon_0}{e}kB_1\left( (1+\beta_\perp)\frac{T_0}{l_T} + \frac{T_0}{l_n} \right).
\label{jdote}   \end{split}
\end{align}
With use of $\nabla.\mathbf{J}=0$, the heat flux terms in eqn. (\ref{heat}) resulting from the current $\mathbf{J}$ are also of similar magnitude to eqn. (\ref{jdote}). In hot weakly coupled plasmas, these terms containing $\mathbf{J}$ are small compared to the thermal heat flux. For example, the linearized contribution from the $\kappa_\wedge$ Righi-Leduc cross gradient heat flux term from eqn. (\ref{heat}) is
\begin{align}
\begin{split}
-\nabla.\mathbf{q}_{\wedge} \simeq \nabla.\left( \frac{n_eT_e\tau\kappa_\wedge}{m_e}\mathbf{\hat b\times \nabla} T_e\right)= \frac{n_0T_0\tau^2e}{m_e^2}\frac{\partial T_0}{\partial x}\kappa_{0\wedge} ikB_1\label{rl}
   \end{split}
\end{align}
The ratio of this term with the first term of eqn. (\ref{jdote}) is
\begin{align}
\begin{split}
\frac{|\nabla.\mathbf{q}_{\wedge}|}{|\mathbf{J.E}|} =  \frac{n_0T_0\tau^2e^2\kappa_{0\wedge}}{m_e^2c^2\epsilon_0 (1+\beta_\perp)} \simeq \Gamma^2.
   \end{split}
\end{align}

 Since we assume a hot weakly coupled plasma with $\Gamma \gg 1$, the terms containing $\mathbf{J}$ in eqn. (\ref{energyeqn}) are also negligible. On neglecting these current terms, the energy equation becomes 
\begin{align}
    \frac{3n_0}{2}\frac{\partial T_1}{\partial t}=\nabla.\left[\frac{n_eT_e\tau}{m_e}\left(\kappa_\perp\nabla T_e + \frac{\kappa_{0\wedge} e\tau}{m_e}\mathbf{B}\times \nabla T_0\right) \right]+S.
\end{align}
There is a diffusive heat flux contribution from both the zeroth-order temperature profile and the transverse smoothing of the temperature perturbation. There is also the divergence of the Righi-Leduc heat flow. Substituting for $T_1(y,t)$ and $B_1(y,t)\propto \exp(\gamma t + iky)$, as well as eqn. (\ref{rl}), leads to the linearized form
\begin{align}
&(\gamma + D_Tk^2) T_1 = \tilde Q  + \frac{2T_0\tau^2e}{3m_e^2}\frac{\partial T_0}{\partial x}\kappa_{0\wedge} ikB_1 + \frac{2S}{3n_0},\label{T1}\\
&\tilde Q = \frac{1}{n_0}\frac{\partial}{\partial x}\left(n_0D_T\frac{d T_0}{d x}\right),\,\,\,\,\,\,\,\,D_T = \frac{2T_0\tau\kappa_\perp}{3m_e}.
\label{linheat}
\end{align}
The zeroth-order heat diffusion term $\tilde Q$ results from the heat flux down the zeroth-order temperature gradient in the $x$ direction. This heat flux causes a rapid relaxation to the steady state zeroth-order temperature profile $T_0(x)$, given by the solution to $\tilde Q=-2S/(3n_0)$. As such, the zeroth-order temperature profile along $x$ is set by the energy source term $S$ and so the $\tilde Q$ and $S$ terms cancel each other. Even if $T_0$ does have a time dependence, $\tilde Q$ will be negligible so long as $kL\gg 1$. Bissell \emph{et al.} \cite{bissell2015nernst} have argued that neglecting $\tilde Q$ is only consistent if we also take $N=0$, as assumed here.

Taking the steady state temperature profile with $\tilde Q=-2S/(3n_0)$, and using the total diffusivity $D=D_R+D_T$, substitution of eqn. (\ref{T1}) into eqn. (\ref{B1}) results in the local dispersion relation
\begin{align}
&\gamma^2 +\gamma Dk^2 + D_TD_Rk^2(k^2-k_c^2)=0,\label{dispersion}\\
&\gamma=-\frac{Dk^2}{2}\pm\sqrt{D_TD_Rk^2(k_c^2-k^2) + \frac{D^2k^4}{4}}\label{growthrate}\\
\label{kc}&k_c^2 = \Gamma^2\frac{\kappa_{0\wedge}}{\kappa_\parallel\alpha_\parallel}\frac{1}{l_T}\left(\frac{1}{l_n} - \frac{d\beta_\parallel}{d \bar Z}\frac{\bar Z}{l_Z}\right),\\
&\Gamma^2=\frac{9\pi}{2}\frac{T_0}{m_ec^2}\left(\frac{\Lambda}{\ln (\Lambda)}\right)^2 =\frac{T_0\tau^2 n_0e^2}{\epsilon_0m_e^2c^2}
\end{align}

There is a root $\gamma>0$ for $0<|k|<k_c$. Instability therefore occurs when $k_c^2>0$. The standard thermomagnetic instability is recovered for $\nabla\bar Z=0$. Since $\kappa_{0\wedge}>0$, instability occurs when $l_nl_T>0$ and the density and temperature gradients have the same direction. 

However, thermomagnetic instability is still possible if $\nabla n_e=0$ but $\nabla \bar Z$ is not. The coefficient derivative $\beta_\parallel'(\bar Z)$ is always positive and takes a maximal value of $\simeq 0.3$ for $\bar Z=1$, typical of hydrogen fusion fuel and many situations in astrophysics. Inspection of eqn. (\ref{kc}) then shows that $l_Tl_Z<0$ is required for instability, such that the temperature and $Z$ gradients are oppositely directed. This naturally arises in an optically thin high-energy-density plasma, since regions of higher ion charge state $\bar Z$ will often have a higher radiative cooling and reach a lower temperature than the surrounding lower $\bar Z$ plasma.

In the limit of low $k$, the growth rate reduces to $\gamma=\sqrt{D_TD_R}k_ck$. Since $k_c\propto\Gamma$, the instability is fastest growing in hot, weakly coupled plasmas, such that $\Gamma\gg 1$. This also justifies neglecting the terms containing $\mathbf{J}$. In addition, the growth rate increases with the gradients in the zeroth-order fluid quantities. However, in conditions where $\lambda_\mathrm{mfp}$ approaches the gradient scale lengths $l_T$ and $l_n$, the calculated growth rate will be reduced by non-local effects \cite{sherlock2020suppression}.

  \begin{figure}[t]
  \includegraphics{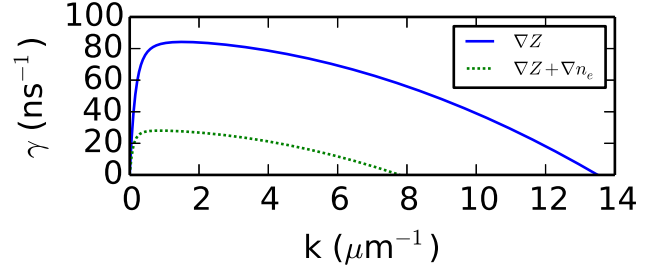}
  \caption{Growth rate of the Z-gradient thermomagnetic instability for the case of $\bar Z=1$, electron density $n_e=2\times10^{25}\,$cm$^{-3}$, $T_e=5\,$keV, and opposed gradients in $T_e$ and $\bar Z$ each with gradient scale-lengths of $3\,\mu$m. The solid line shows eqn. (\ref{growthrate}) for $\nabla n_e=0$. The dotted line shows the more realistic case for fusion fuel contaminants, with an additional $\nabla n_e$ towards the higher $\bar Z$ region with $l_n=15\,\mu$m. The growth rate and cutoff wavenumber are reduced due to the counter-acting Biermann source term. This plot neglects the Nernst term $N$ and the fluid motion $\mathbf{u}$.}
   \label{growth rate}
\end{figure}

Neglecting the Nernst advection term $N$ and the zeroth-order thermal smoothing $\tilde Q$, the growth rate (\ref{growthrate}) is plotted in Fig. \ref{growth rate} for the case of $\bar Z=1$, $n_e=2\times10^{25}\,$cm$^{-3}$, $T_e=5\,$keV and $l_Z=-l_T=3\,\mu$m. These parameters are the expected values for the higher-Z fill tube contaminant jet that can enter inertial confinement fusion fuel \cite{weber2020mixing}. The solid line shows the rather contrived case of $\nabla n_e=0$, such that the standard thermomagnetic instability is eliminated. The growth rate initially increases with $k$, before becoming damped by the resistive and thermal diffusion. This dissipation eventually leads to a wavenumber cutoff at $k=k_c$. The peak growth rate is $\gamma\simeq 80\,$ns$^{-1}$. This means there could be time for many e-foldings within the $100\,$ps fusion fuel stagnation period. 

An additional case of interest is that of plasma with uniform total pressure $P$. This is especially pertinent to our assumption that $\mathbf{u}=0$, requiring $\nabla P=0$ for continued validity on hydrodynamic timescales. In general, this case will not have $\nabla n_e=0$, and so the process will be some combination of the $Z$-gradient instability and the standard Tidman-Shanny thermomagnetic instability \cite{tidman1974field}. Depending on the sign of $\nabla n_e\cdot\nabla\bar Z$, the two source terms will either reinforce each other or cancel out. Fig. \ref{growth rate} shows an additional case with $\nabla n_e\cdot\nabla \bar Z > 0$, in which the Biermann term counteracts and partially stabilizes the $\nabla \bar Z$ source term. This is closer to the true conditions of the fusion fuel contaminant jet, since they usually evolve in pressure equilibrium with the surrounding fuel \cite{weber2020mixing}. In these more realistic conditions, the growth rate is either reduced or fully stabilized by the Biermann term. 

It is important to note that the original Tidman-Shanny derivation requires $\nabla n_e\cdot\nabla T_e>0$ for the standard instability, typically requiring that $\nabla P\neq 0$, for example in a laser ablation front. However, with inclusion of the $Z$-gradient source term in multi-species plasma, instability can be achieved even when $\nabla P=0$. This means growth could even occur in isobaric plasma, over periods much longer than the hydrodynamic time-scale.

\section{Numerical simulations and Nernst stabilization}
Referring back to the induction equation (\ref{induction}), the linearized analysis is identical to that of the standard thermomagnetic instability, except for the replacement $\nabla n_e \rightarrow \nabla n_e - n_e\beta_\parallel'(\bar Z)\nabla\bar Z$, leading to the additional $\bar Z$-gradient source term. As such, the existing linear theory framework can be generalized to the case of non-zero $\bar Z$ gradients. In particular, the discussion of non-zero Nernst advection $N$ \cite{bissell2015nernst, sherlock2020suppression, hirao1981magnetic, urpin2019nernst}, fluid motion $\mathbf{u}$ \cite{ogasawara1980hydrodynamic} and greater magnetization \cite{fruchtman1992thermomagnetic} will also apply to unstable $\bar Z$ gradients. Positive Nernst term $N$ can enhance the instability and lead to $k_c^2>0$, even when $l_nl_T<0$ or $l_Zl_T>0$. However, Bissell \emph{et al.} \cite{bissell2015nernst, bissell2013super} have argued that this does not constitute a true instability since the growth is due to conservative Nernst compression of the existing field, rather than unstable coupling of self-generated field with the magnetized heat flux. The case of $N<0$ will always advect away the magnetic field and dampen its growth. Due to its origin also in the zeroth-order heat flux, the magnitude of $N$ is usually similar to $\gamma B_1$. This means that the Nernst term is likely to advect the self-generated field away from the unstable region on a similar time-scale to the instability growth. The Nernst term is also dependent on the $x$ derivative, meaning the full two-dimensional $B_1(x,y, t)$ profile must be considered. 

 To this end, the growth rate (\ref{growthrate}) will now be verified via a two-dimensional single-fluid extended magneto-hydrodynamic simulation. This will also highlight the stabilizing effect of the Nernst advection. The magnetic field was evolved via eqn. (\ref{induction}) and the electron temperature was evolved via eqn. (\ref{heat}) and eqn. (\ref{energyeqn}). In addition, to isolate the stabilization mechanism, fluid motion $\mathbf{u}$ was set to zero. The Coulomb logarithm used a fixed value of $4$.

We used a two-dimensional Cartesian periodic domain, with side lengths $L_x=4\,\mu$m and $L_y=1\,\mu$m, uniform grid resolution $25\,$nm and a time-step of $0.2\,$fs. The simulation was initialized to depict a carbon contaminant jet entering a dense inertial confinement fusion hot-spot. To eliminate the Biermann battery and therefore isolate the Z-gradient instability, the electron density was uniform at $n_e=2\times10^{25}\,$cm$^{-3}$. Deuterium (Z=1) and Carbon (Z=6) species were initialized with number density profiles $n_H = n_e(0.8 - 0.2\cos(2\pi x/L_x))$ and $n_C=(n_e-n_H)/6$. The electron temperature profile was initialized as uniform with a small transverse perturbation $T_e=T_0(1.0+0.001\cos(2\pi y/L_y))$, with $T_0=4\,$keV. A zeroth-order temperature profile in the x-direction was created using a fixed spatially dependent energy source term $S(x)=-S_0\cos(2\pi x/L_x)$, with $S_0=2\times 10^{22}\,$Wcm$^{-3}$, loosely representing an increased radiative loss in the carbon region around $x=0$. Since the average value of $S$ across the domain is zero, it serves to simply redistribute the plasma internal energy in the $x$ direction and maintain a zeroth-order temperature profile with the required gradients in the $x$ direction. This source term was active throughout the simulation, acting to maintain the temperature profile in the $x$ direction in the presence of the smoothing via heat conduction. 

The transport coefficients used the fit functions presented in ref. \cite{epperlein1986plasma}. Due to issues with the $\beta_\perp$ coefficient fit function presented in ref. \cite{epperlein1986plasma} at low magnetization, we simply used $\beta_\perp=\beta_\parallel$ in the simulations. This approximation should be accurate for these test cases with $\chi\ll 1$. This issue with the results of ref. \cite{epperlein1986plasma}, which leads to a huge over-estimation of the cross-gradient Nernst advection, will be discussed in a future publication.

\begin{figure}[t!]
  \includegraphics{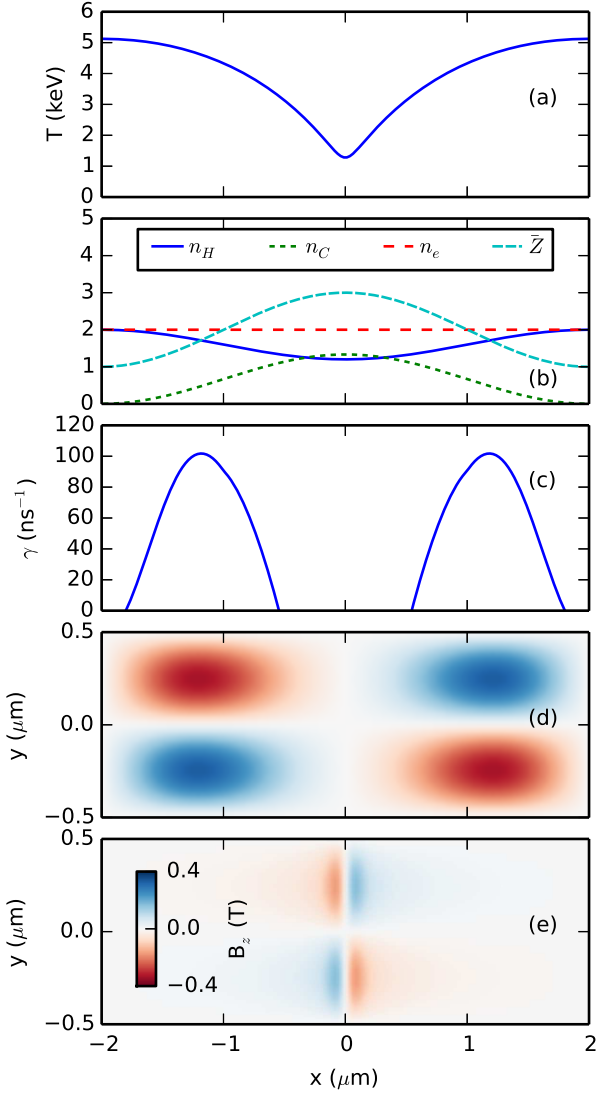}
  \caption{Results of the two-dimensional Cartesian extended magneto-hydrodynamics simulation of the Z-gradient instability. Plasma was initialized with uniform electron density $n_e=2\times 10^{25}\,$cm$^{-3}$ and ions were kept static with zero fluid velocity. (a) Line-out at $y=0$ of the $x$ direction temperature profile after $6\,$ps. This has reached steady state via the action of the fixed energy source term $S(x)=-S_0\cos(2\pi x/L_x)$. There was also an initial sinusoidal transverse temperature perturbation of wavelength $1\,\mu$m. (b) Plots of the fixed Deuterium $n_H$, Carbon $n_C$ and electron $n_e$ species number densities, as well as the average ion charge state $\bar Z$. $n_H$ and $n_e$ are normalized to $10^{25}\,$cm$^{-3}$ and $n_C$ is normalized to $10^{24}\,$cm$^{-3}$. (c) The resulting local growth rate of the $\bar Z$ gradient instability, excluding the Nernst effect, as calculated using eqn. (\ref{growthrate}). (d) The two-dimensional magnetic field profile after $6\,$ps, for the case without Nernst advection. (e) The two-dimensional magnetic field profile after $6\,$ps, for the case with Nernst advection.}
   \label{figsim}
\end{figure}

Equations (\ref{induction}) and (\ref{energyeqn}) were numerically integrated for $50\,$ps using a second order central finite difference explicit Runge-Kutta method, starting from $\mathbf{B}=0$. The source term $S$ resulted in a steady state zeroth-order temperature profile within a few picoseconds. The source term $S$ removes energy from the central region around $x=0$ and transfers it to the region around $x=\pm 2\,\mu$m. This effect is balanced by the electron heat flux towards $x=0$, resulting in a steady-state zeroth-order temperature profile in the $x$ direction with an imposed initial perturbation in the $y$ direction. Fig. \ref{figsim}a shows the resulting well established temperature profile $T_0(x)$ after $6\,$ps. Although $S$ has a sinusoidal profile, the resulting $T_0$ profile is not sinusoidal due to the spatial variation of the heat conductivity. Fig. \ref{figsim}b shows the fixed ion and electron number density profiles, with the carbon located in the central region. It also shows the resulting average ion charge state $\bar Z$, which has an oppositely directed gradient to the temperature gradient shown in Fig. \ref{figsim}a. This is the condition for instability according to eqn. (\ref{growthrate}). These profiles result in $\Gamma\simeq 30$ in the hotter regions.

Fig. \ref{figsim}c shows the resulting growth rate, as calculated using the profiles in Figs. \ref{figsim}a,b and eqn. (\ref{growthrate}). It is positive in the hotter plasma regions with the steepest gradients. The growth rate is negative in the central, colder and more resistive plasma. This is because the initial single mode transverse perturbation exceeds the cutoff wavenumber $k_c$. The growth is also stabilized in the edge regions, as the gradients are too shallow and so again $k>k_c$.

The simulation was repeated with the Nernst advection disabled. Fig. \ref{figsim}d and Fig. \ref{figsim}e show the two-dimensional magnetic field profiles after $6\,$ps for the cases without and with Nernst advection. The magnetic growth is stronger in the case without Nernst advection in Fig. \ref{figsim}d. It is also clear that the growth rate is fastest in the hotter regions, despite the shallower temperature gradient here. The overall magnetic profile agrees well with the prediction in Fig. \ref{figsim}c, although there are some differences due to minor terms that were neglected in the local growth rate derivation.

When the nonlinear Nernst advection is included, as shown in Fig. \ref{figsim}e, it is clearly a dominant term and it acts to stabilize the instability. This is because any field that arises in the unstable region is quickly advected into the central colder region by the Nernst term, and is then dissipated by the greater resistivity here.  

\begin{figure}[t!]
  \includegraphics{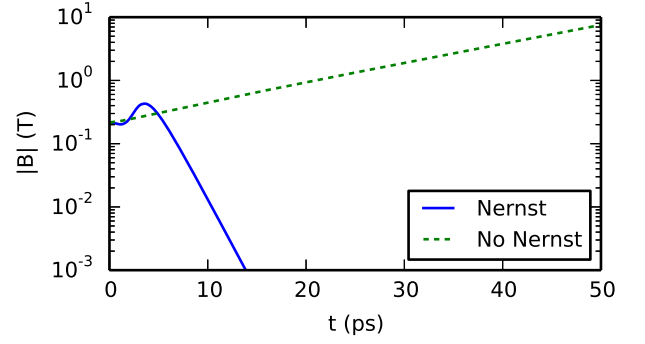}
  \caption{ The maximal magnetic field magnitude as a function of time, for the simulations with and without Nernst advection. }
   \label{figsimt}
\end{figure}
Fig. \ref{figsimt} shows the resulting peak magnetic field as a function of time, starting from $B_1(t=0)=0$. A magnetic field of magnitude $\simeq 0.2\,$T quickly arises in both cases due to the initial transverse temperature perturbation and the Z-gradient source term. In the case without Nernst advection, exponential growth occurs with growth rate $\simeq75\,$ns$^{-1}$, in approximate agreement with the peak growth rate of $100\,$ns$^{-1}$ given in Fig. \ref{figsim}c. The discrepancy could be due to the fact that the simulation only has $kL\simeq 15$, meaning the local approximation is questionable. Thermal smoothing of the perturbation in the $x$-direction will be significant, whereas the local growth rate derivation did not consider this.

It should be noted that, in Fig. \ref{figsimt}, for a brief time up until $t=4\,$ps, Nernst compression of the transient magnetic field actually increases the field magnitude above that of the unstable case without the Nernst advection. This is in agreement with the linear thermomagnetic growth rates including the Nernst advection term \cite{hirao1981magnetic}, which can take on values higher than eqn. (\ref{growthrate}). However, the later nonlinear stage of the Nernst advection serves to simply compress the magnetic field into the colder region, where the growth rate in Fig. \ref{figsim}c is negative. The compressed magnetic field region reaches an equilibrium size in the $x$ direction, shown in Fig. \ref{figsim}e, in which the Nernst compression towards $x=0$ is balanced by the resistive diffusion of the magnetic field. The resistive diffusion between the positive and negative regions of $B_z$ also reduces the maximal field magnitude over time. The temperature perturbation is also dissipated by thermal conduction, meaning the $B_1$ and $T_1$ transverse perturbations decay exponentially, shown in Fig. \ref{figsimt}. 

Other simulations with various initial parameters and profiles exhibited the same qualitative behavior, with negative growth rate in the central colder region. This suggests that the global Nernst term $N$ is effective in stabilizing the thermomagnetic growth, even in conditions where the local growth rate eqn. (\ref{growthrate}) is positive in some regions. This remained true when the initial density, temperature and perturbation wavelength were varied by a factor of 100, and the boundary profiles were steepened by varying the profile of the source term $S$. Similar qualitative behavior was also seen for cases with $\nabla n_e\neq 0$. The Nernst stabilization was also still dominant in simulations with $|l_Z|\ll |l_T|$. 

This conclusion is in agreement with the numerical results of reference \cite{sherlock2020suppression}, in which  Nernst advection, among other things, stabilized the standard thermomagnetic instability. Inclusion of ion hydrodynamics may change this conclusion, although in typical conditions at laser ablations fronts the Nernst velocity greatly exceeds the fluid velocity, meaning the Nernst stabilization is likely to remain the dominant effect. The Nernst velocity is also of similar magnitude to the fluid velocity in stagnated inertial confinement fusion fuel \cite{walsh2017self}.

\section{Summary}

In summary, we have reported a field generating thermomagnetic instability mechanism that arises from the coupling of the thermal force Z-gradient magnetic source term with the deflected Righi-Leduc heat flow. As such, the instability relies on a collisional mechanism that is fundamentally different to the standard thermomagnetic instability. Unstable conditions rely on opposing gradients in ion charge state and electron temperature, as are found around fusion fuel contaminant regions. The linearized growth rate is similar to that of the standard thermomagnetic instability \cite{tidman1974field, bolshov1974spontaneous}. However, in agreement with later numerical work \cite{sherlock2020suppression} on the standard instability, we find that the Z-gradient instability is also stabilized by Nernst advection across a wide range of conditions. Magnetic field is quickly advected out of the unstable region and dissipated. This is because the Nernst advection, with its physical origin also due to the electron heat flux, is always of similar magnitude to the linearized thermomagnetic instability growth rate. The global nature of this Nernst advection means that multi-dimensional ExMHD simulations are required to assess the stability of any given configuration, also considering the added effects of fluid motion and radiation transport. These global effects cannot be included in a simple local analytic growth rate.

\section*{Authors' contributions}
J.S. conducted the analysis and wrote the paper. H.L. participated in the discussion and interpretation of results. 
\begin{acknowledgments}
We gratefully acknowledge the support of the U.S. Department of Energy through Los Alamos National Laboratory under Laboratory Directed Research and Development project number 20180040DR and the Center for Nonlinear Studies.  The authors wish to thank the scientific computing staff at Los Alamos National Laboratory. 
\end{acknowledgments}

\section*{Data Availability}
The data that support the findings of this study are available on request from the corresponding author. The data are not publicly available due export control restrictions at Los Alamos National Laboratory.


%

\end{document}